%% file: main.tex
\documentclass[runningheads,a4paper]{llncs}

\usepackage{amssymb}
\usepackage{graphicx}
\usepackage{subfigure}
\usepackage{tabularx}
\usepackage{listings}
\usepackage[english]{babel}
\usepackage{times}
\usepackage{AMMALanguages}
\usepackage{multicol}
\usepackage{caption}

\usepackage{url}

\newcommand{\keywords}[1]{\par\addvspace\baselineskip
\noindent\keywordname\enspace\ignorespaces#1}

\definecolor{commentcolor}{RGB}{78,156,93}

\input{commentSupport}

\begin{document}

\mainmatter  

\title{Towards the Automation of Metamorphic Testing in Model Transformations}

\titlerunning{Towards the Automation of Metamorphic Testing in Model Transformations}

%
%
\author{Javier Troya \and Sergio Segura \and Antonio Ruiz-Cort\'{e}s}
\authorrunning{Towards the Automation of Metamorphic Testing in Model Transformations}

\institute{Department of Computer Languages and Systems\\
Universidad de Sevilla, Spain\\
\{jtroya, sergiosegura, aruiz\}@us.es}

%
%

\toctitle{Towards the Automation of Metamorphic Testing in Model Transformations}
\tocauthor{Troya, Segura, Ruiz-Cort\'{e}s}
\maketitle

\vspace{-5pt}

\begin{abstract}
Model transformations are the cornerstone of Model-Driven Engineering, and provide the essential mechanisms for manipulating and transforming models.
Checking whether the output of a model transformation is correct is a manual and error-prone task, this is referred to as the oracle problem in the software testing literature.
The correctness of the model transformation program is crucial for the proper generation of its output, so it should be tested.
Metamorphic testing is a testing technique to alleviate the oracle problem consisting on exploiting the relations between different inputs and outputs of the program under test, so-called metamorphic relations.
In this paper we give an insight into our approach to generically define metamorphic relations for model transformations, which can be automatically instantiated given any specific model transformation.

\end{abstract}

\vspace{-20pt}

\section{Introduction}
\label{sec:Introduction}
\vspace{-5pt}
\noindent \input{sections/01_Introduction}
\vspace{-10pt}

\section{Approach}
\label{sec:Approach}
\vspace{-10pt}
\noindent \input{sections/02_Approach}

\vspace{-10pt}
\section{Next Steps and Observations}
\label{sec:Conclusion}
\vspace{-5pt}
\noindent \input{sections/03_Conclusion}

\vspace{-5pt}

\subsubsection*{Acknowledgments.} This work has been partially funded by the European Commission (FEDER) and Spanish Gov. under CICYT project BELI (TIN2015-70560-R), and by the Andalusian Gov. projects THEOS (TIC-5906) and COPAS (P12- TIC-1867).

\bibliographystyle{splncs03}
\bibliography{literature}

\end{document}

%% file: commentSupport.tex
\usepackage[normalem]{ulem} 
\usepackage{xcolor}
\definecolor{colgray}{gray}{0.75}

\usepackage{ifthen}
\usepackage{amssymb}
\newboolean{showcomments}
\setboolean{showcomments}{true} 
\ifthenelse{\boolean{showcomments}}
  {\newcommand{\nb}[2]{
    \fcolorbox{gray}{yellow}{\bfseries\sffamily\scriptsize#1}
    {$\blacktriangleright$#2$\blacktriangleleft$}
   }
   
  }
  {\newcommand{\nb}[2]{}
   
  }

%% file: sections/01_Introduction.tex
Model Transformations (MTs) are the cornerstone of Model-Driven Engineering (MDE).
They provide the essential mechanisms for manipulating and transforming models.
Checking whether the output of a model transformation is correct is a manual and error-prone task, this is refereed to as the oracle problem in the software testing literature.
Indeed, the quality of the generated software artifacts is highly affected by the correctness of the developed model transformations.
For this reason, several approaches have been proposed that verify the correct behavior of the transformations using formal methods~\cite{Troya11,AmraniLSCDVTC12} or certify their behavior for a selected set of test models mainly to identity bugs in a cost-effective way~\cite{GV2011,VallecilloGBWH12}.

Metamorphic Testing (MT')~\cite{Segura16} is a methodology designed to alleviate the oracle problem.
Different from conventional testing strategies, MT' consists on exploiting the relations between different inputs and outputs of the program under test, so-called Metamorphic Relations (MRs).
In practice, MRs define possible modifications to a test input and how those changes are propagated to the program output.
A basic example of MR can be defined for the program that computes the sine function. 
Let us suppose we want to know the exact value of $\sin(5)$.
Is an observed output of $0.091$ correct?
A mathematical property of the sine function states that $\sin(x) = \sin(\pi-x)$, and we can use this to test whether $\sin(5) = \sin(\pi-5)$ without knowing the concrete values of either sine calculation.
Currently, the biggest limitation of MT' has to do with the definition of the MRs.
In fact, the automatic generation of MRs has been acknowledged as one of the big challenges in MT'~\cite{Segura16}.

\begin{figure}[t]%
	\centering%
	\includegraphics[width=.7\columnwidth]{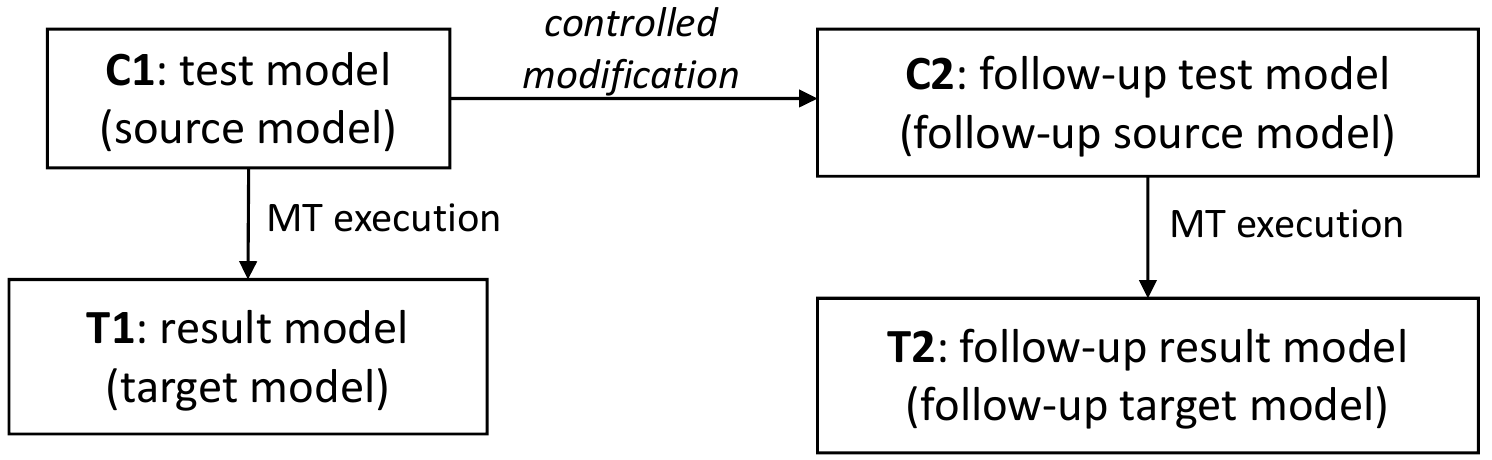}%
	\caption{Metamorphic Testing in Model Transformations.}%
	\label{fig:MTinMT}%
\end{figure}

Figure~\ref{fig:MTinMT} displays the scenario of MT' in MTs.
We have a test model (C1), which is typically the source model.
If we apply the MT, we obtain the result model (T1), i.e., the target model.
By applying a controlled modification in the test model, we obtain the follow-up test model (C2).
If we now apply the same MT to C2, we get the follow-up result model (T2).
In this context, a MR considers the modification done in C2 with respect to C1, and the consequences that this has in T2 with respect to T1.
We show an example in next section.

As far as we are concerned, there is only one approach that applies MT' in MTs~\cite{SEKE14}.
The authors demonstrate the effectiveness and feasibility of its application, although they apply it manually in a specific scenario, for which they define the MRs. In our approach, we propose to automatically generate MRs for any model transformation, as we explain in the next section. Then, in Section~\ref{sec:Conclusion} we describe our next steps. 

%% file: sections/02_Approach.tex
The goal of our approach is to automate the process of metamorphic testing (MT') in model transformations (MTs).
Thereby, we propose the automatic generation of metamorphic relations (MRs) for any model transformation.
We work with transformations written in the ATL Transformation Language due to its importance both in academia and industry.

\begin{figure}[t]
\centering
\subfigure[Simple Generic Trace.]{
\includegraphics[width=.7\textwidth]{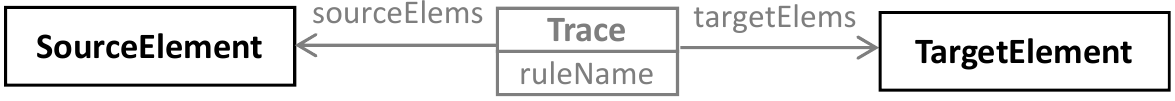}
\label{fig:genericTrace}
}
\subfigure[Traces for the transformation excerpt of Listing~\ref{lst:class2relational}.]{
\includegraphics[width=.8\textwidth]{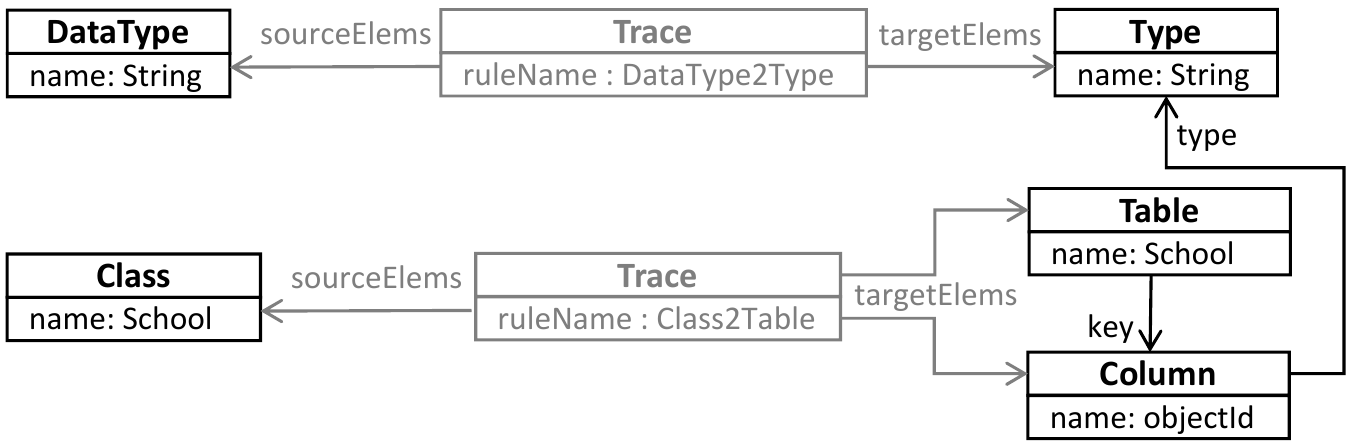}
\label{fig:class2relationalTrace}
}
\label{fig:traces}
\caption{Generic and instantiated traces}
\end{figure}

In order to automatically extract information out of a transformation, we make use of explicit trace models.
A trace model can be automatically obtained from a transformation execution, e.g., by using Jouault's \emph{TraceAdder}~\cite{Jouault05}, and is composed of a set of traces, one for each rule execution.
A trace captures the name of the applied rule and the elements of the source model (\emph{sourceElems} relationship) that are used to create new elements in the target model (\emph{targetElems} relationship).
Therefore, by navigating the trace model, we know which target element(s) have been created from which source element(s) and by which rule.
A simple example of a generic trace is shown in Figure~\ref{fig:genericTrace}.
Please note that more than one element may appear as \emph{sourceElems} and \emph{targetElems}.
We consider this trace as generic because each of the three elements appearing in it (\emph{SourceElement}, \emph{Trace} and \emph{TargetElement}) can be instantiated in a particular scenario.


The idea of our approach is to define generic MRs for generic traces.
These MRs can then be instantiated together with the generic traces.
For instance, considering the generic trace of Figure~\ref{fig:genericTrace}, we know that if we have a test model (C1, Figure~\ref{fig:MTinMT}) and we add an element of type \emph{SourceElement} in the follow-up test model (C2), then an element of type \emph{TargetElement} is created for it in the follow-up result model (T2).
This means that T2 has one more element of this type than T1.
Having this information into account, we can define the first generic MR shown in Listing~\ref{lst:genericMR}, written in the OCL language.
Besides, the number of elements of any other type should remain the same in T1 and T2, so further MRs can be defined, such as the second one in the same listing.

\begin{lstlisting}[language=ATL, breaklines=true, caption={Generic MRs for the addition of a \emph{SourceElement}},label={lst:genericMR}]
T1_TargetElement.allInstances()->size()=T2_TargetElement.allInstances()->size()-1
T1_AnyOtherType.allInstances()->size()=T2_AnyOtherType.allInstances()->size()
\end{lstlisting}

%

In order to show an example of instantiation of the generic MRs shown before, we choose the well-known \emph{Class2Relational} case study.
We will focus on the excerpt of the transformation, which has been slightly modified for simplicity in the explanation, shown in Listing~\ref{lst:class2relational}.
If we have a source model with a \emph{DataType} and a \emph{Class} and we apply the transformation, the resulting trace model is depicted in Figure~\ref{fig:class2relationalTrace}.
We can see a trace that reflects the creation of a \emph{Type} from a \emph{DataType} and another one that stores the creation of a \emph{Table} and a \emph{Column} from a \emph{Class}.


\begin{multicols}{2}[\captionof{lstlisting}{Excerpt of Class2Relational transformation}]
\begin{lstlisting}[language=ATL,label={lst:class2relational}]
rule DataType2Type {
  from
    dt : Class!DataType
  to
    out : Relational!Type (
      name <- dt.name
    )
}

rule Class2Table {
  from c : Class!Class
  to
    out : Relational!Table (
      name <- c.name,
      key <- key),
    key : Relational!Column (
      name <- 'objectId',
      type <- thisModule.objectIdType)
}
\end{lstlisting}
\end{multicols}

\vspace{-10pt}

Since these two traces are instantiations of the generic one shown in Figure~\ref{fig:genericTrace}, we can also instantiate the MRs shown in Listing~\ref{lst:genericMR}. 
In particular, we have two scenarios.
The first one consists of adding an element of type \emph{DataType} in C2, what yields the MRs shown in Listing~\ref{lst:class2relationalMR}.
In the second scenario we add an element of type \emph{Class} in C2, obtaining the MRs shown in Listing~\ref{lst:class2relationalMR2}.

\begin{lstlisting}[language=ATL, breaklines=true, caption={MRs for the addition of a \emph{DataType} in C2},label={lst:class2relationalMR}]
T1_Type.allInstances()->size()=T2_Type.allInstances()->size()-1
T1_Column.allInstances()->size()=T2_Column.allInstances()->size()
T1_Table.allInstances()->size()=T2_Table.allInstances()->size()
\end{lstlisting}

\begin{lstlisting}[language=ATL, breaklines=true, caption={MRs for the addition of a \emph{Class} in C2},label={lst:class2relationalMR2}]
T1_Column.allInstances()->size()=T2_Column.allInstances()->size()-1
T1_Table.allInstances()->size()=T2_Table.allInstances()->size()-1
T1_Type.allInstances()->size()=T2_Type.allInstances()->size()
\end{lstlisting} 

%% file: sections/03_Conclusion.tex
In this paper we have given an insight into our approach to automate the generation of MRs for MTs.
We identify generic patterns in the traces, from which we define generic MRs organized as well in patterns.
For instance, one pattern is the generic trace and MRs we have shown in this paper.
Despite its simplicity, we are performing ongoing works defining more patterns where elements, attributes and relationships are taken into account, so that we end up with a large set of MRs.
One of the purposes of the generated MRs is to automate regression tests, since they can be checked as to whether they hold in different versions of the model transformation program and for any test model.

As mentioned, our approach takes as input one or more executions of a MT, i.e., the resulting trace models.
The number of executions of the MT received and their size influence the completeness of the MRs generated.
For instance, if a rule is never applied in any of the executions received as input, no MRs will consider its behavior.